\documentclass[twocolumn,showpacs,amsmath,amssymb,10pt,aps,superscriptaddress]{revtex4}
\usepackage{graphicx,color}
\usepackage{amsmath}
\usepackage{amssymb}
\usepackage{dsfont}
\usepackage{bbm}
\usepackage{color}
\usepackage{amsfonts}
\usepackage{bm}
\usepackage[utf8]{inputenc}
\def\oper{{\mathchoice{\rm 1\mskip-4mu l}{\rm 1\mskip-4mu l}
{\rm 1\mskip-4.5mu l}{\rm 1\mskip-5mu l}}}
\def\<{\langle}
\def\>{\rangle}
\newtheorem{theorem}{Theorem}
\newtheorem{Lemma}{Lemma}

\newtheorem{Cor}{Corollary}
\newtheorem{Def}{Definition}
\newtheorem{Remark}{Remark}
\newtheorem{Example}{Example}
\newtheorem{Proposition}{Proposition}

\newcommand{\LH}{{{\rm L}}(\mathcal{H})}
\usepackage{color}
\usepackage{soul}

\makeatletter

\begin{document}

\title{Construction of propagators for divisible dynamical maps}


\author{Ujan Chakraborty}

\affiliation{{Department of Physical Sciences, Indian Institute of Science Education and Research Kolkata,
		Mohanpur, Nadia - 741246, West Bengal, India}}

\author{Dariusz Chru\'{s}ci\'{n}ski}

\affiliation{Institute of Physics, Faculty of Physics, Astronomy and Informatics,
Nicolaus Copernicus University, Grudziadzka 5/7, 87-100 Toru\'{n},
Poland}

\begin{abstract}
Divisible dynamical {maps} play {an} important role in characterizing Markovianity on the level of quantum evolution. Divisible maps provide important generalization of Markovian semigroups. Usually one analyzes either completely positive or just positive divisibility meaning that the corresponding propagators are defined in terms of completely positive or positive maps, respectively. For maps which are invertible at any moment of time the very existence of propagator is already guaranteed and hence the only issue is (complete) positivity and trace-preservation. However, for maps which are not invertible the problem is much more involved since even the existence of a propagator is not guaranteed. In this paper we propose a simple method to construct propagators of dynamical maps using the concept of generalized inverse. We analyze both time-continuous and time-discrete maps. Since the generalized inverse is not uniquely defined the same applies for the corresponding propagator. {In simple examples of qubit evolution we analyze it turns out that the additional requirement of complete positivity possibly makes the propagator unique.}
\end{abstract}

\maketitle

\section{Introduction}

The evolution of open quantum systems \cite{Open1,Open2} attracts a lot attention due to
rapidly developing modern fields of science and quantum technologies like quantum communication or quantum computation \cite{QIT}.
Any quantum evolution is represented by a {\em quantum dynamical map}, that is, a continuous family of linear maps $\{\Lambda_t\}_{t\geq 0}$

\begin{equation}\label{}
  \Lambda_t : \LH \to \LH ,\ \ (t\geq 0) ,
\end{equation}
where $\LH$ denotes the space of linear operators acting on the system's Hilbert space $\mathcal{H}$ (considered to be finite dimensional, in this paper).  One requires that for any $t\geq 0$ the map $\Lambda_t$ is completely positive and trace-preserving (CPTP) \cite{Paulsen,Stormer,Wolf}. Moreover, it satisfies the standard initial condition $\Lambda_{t=0}={\rm id}$, where ${\rm id}$ denotes the identity map in $\LH$. Hence, for any $t \geq 0$ the map $\Lambda_t$ represents a quantum channel, a basic object of quantum information \cite{QIT,Wolf}.

Very often one also considers a different scenario: time-discrete dynamical map $\{\Lambda_n\}_{n \geq 0}$, where $n=0,1,\ldots$, and $\Lambda_n : \mathrm{L}(\mathcal{H}_0) \to \mathrm{L}(\mathcal{H}_n)$, where $\mathcal{H}_n$ is a Hilbert space at `time' $n$.  In principle for different $n$ the corresponding Hilbert space may be different. Such scenario is important to deal with many processes like encodings, decodings, quantum measurement, and many others. In this paper we consider both time-continuous and time-discrete scenarios.

In the time-continuous case it well known that under appropriate Markovian approximation the evolution of an open quantum system
can be described by quantum dynamical semigroup satisfying the following master equation

\begin{equation}\label{ME}
  \dot{\Lambda}_t = \mathcal{L} \Lambda_t \ , \ \ \Lambda_0 = {\rm id} ,
\end{equation}
where $\mathcal{L} : \LH \to \LH$ has the celebrated Gorini-Kossakowski-Lindblad-Sudarshan (GKLS) structure \cite{GKS,L}

\begin{equation}\label{}
  \mathcal{L}(\rho) = -i[H,\rho] + \sum_\alpha \gamma_\alpha \left( L_\alpha \rho L_\alpha^\dagger - \frac{1}{2} \{ L^\dagger_\alpha L_\alpha,\rho\} \right) ,
\end{equation}
with Hermitian $H^\dagger = H \in \LH$, arbitrary noise (Lindblad) operators $L_\alpha \in \LH$, and dissipation/decoherence rates $\gamma_\alpha \geq 0$ (we keep $\hbar=1$). Markovian approximation leading to (\ref{ME}) assumes weak coupling between system and its environment and separation of characteristic time scales. Such approximation works well in quantum optical systems where the coupling between atoms/molecules and electromagnetic field is we{a}k. Recently, the notion of non-Markovian quantum evolution received considerable attention (see
review papers \cite{NM1,NM2,NM3,NM4,Piilo}). There is no unique universal approach to deal with quantum (non)Markovianity.  In this paper we concentrate on two very popular approaches based on divisibility of dynamical maps \cite{RHP} and so called information flow \cite{BLP}.
The very concept of divisibility of quantum channels was initiated in \cite{Wolf-1,Wolf-2} and recently reviewed in \cite{Mario} (see also recent eprint \cite{Monachium}). According to \cite{RHP} the evolution is Markovian iff there exist a family of CPTP propagators $V_{t,s}$ $(t\geq s)$ such that $\Lambda_t = V_{t,s} \Lambda_s$ (see Section II for more details). One calls such evolution CP-divisible. Requiring that $V_{t,s}$ defines positive (and not necessarily completely positive map) one calls $\{\Lambda_t\}_{t\geq 0}$ P-divisible (for the intricate relation between P- and CP-divisibility cf. \cite{Fabio}).  Actually, one may introduce a whole hierarchy of so called $k$-divisibility, where $k$ runs from $k=1$ (P-divisibility) up to $d={\rm dim}\, \mathcal{H}$ (CP-divisibility) \cite{Sabrina}.  In a recent paper \cite{Farrukh} so called KS-divisibility was analyzed based on so called Kadison-Schwarz property of the propagator $V_{t,s}$ which is stronger than P- but weaker than CP-divisibility. An interesting relation of divisibility of quantum dynamical maps and collision models was studied in \cite{Sergey-2017}. Moreover $k$-divisibility was linked to  discrimination of quantum channels \cite{Junu-DC}. For recent review of various properties of quantum open systems related to divisibility of dynamical maps see recent review \cite{Chem}.  On the other hand authors of \cite{BLP} proposed the following approach: they call quantum evolution represented by $\{\Lambda_t\}_{t\geq 0}$ Markovian if for any pair of states $\rho_1$ and $\rho_2$ so called information flow

\begin{equation}\label{IF}
  \frac{d}{dt}\| \Lambda_t(\rho_1-\rho_2)\|_1 \leq 0 ,
\end{equation}
where $\|X\|_1$ denotes the trace norm. Since $\| \Lambda_t(\rho_1-\rho_2)\|_1$ corresponds to distingushability of time evolved states at time $t$ inequality (\ref{IF}) states that distinguishability monotonically decreases in time. It is interpreted as a flow of information from the system to the environment. Violation of (\ref{IF}) is therefore interpreted as an information back-flow and provides clear evidence of memory effects and hence non-Markovianity \cite{NM1,NM2,NM3,NM4}. These two approaches are not independent \cite{Angel}. Actually, CP-divisibility implies (\ref{IF}). However, violation of (\ref{IF}) is immediately recognized as a clear sign of non-Markovianity. The approach based on information flow is very popular in the literature.  Intriguing connections of divisibility, information back-flow and quantum correlations were reported recently in \cite{Janek,Acin-cor,Johansson}. We stress, however, that there are other approaches to quantum non-Markovianity (see detailed review in \cite{NM4} and the recent papers  \cite{Modi-1,Modi-2} relating (non)Markovianity and quantum stochastic processes).

In this paper we analyze the problem of construction of propagators $V_{t,s}$ (or $V_{j,i}$ in the time-discrete case). If the map $\Lambda_t$ is invertible for all $t \geq 0$, then $V_{t,s} = \Lambda_t \Lambda_s^{-1}$. For maps which are not invertible this scheme does not work (see also discussion in \cite{Cresser1,Cresser2}). Here we propose a very simple approach based on the concept of generalized inverse {(detailed in Section III)} which reduces to the standard inverse if the map is invertible. Generalized inverse and hence the corresponding propagator {(constructed in section IV)} is not uniquely defined. {We illustrate our construction a simple but instructive example (through sections V and VI). In particular, we study divisible qubit dynamical maps exhaustively (in sections VII and VIII), and} provide several examples of qubit evolution {(in section IX), for which } the non-uniqueness of the propagator is removed when one requires that the propagator is completely positive and trace-preserving.

\section{Divisible  dynamical maps}

\subsection{Time-continuous maps}

Consider a dynamical map $\left\{ \Lambda_{t}\right\} _{t\geq0}$ acting on $\LH$. We assume that the family of maps $\left\{ \Lambda_{t}\right\} _{t\geq0}$ is differentiable (w.r.t. $t$), and ${\rm dim}\, \mathcal{H}=d < \infty$.

\begin{Def}
A quantum dynamical map $\left\{ \Lambda_{t}\right\}_{t\geq0}$ is called divisible if for any $t \geq s$

\begin{equation}\label{DIV}
  \Lambda_t = V_{t,s} \Lambda_s ,
\end{equation}
and $V_{t,s} : \LH \to \LH$. $\{\Lambda_t\}_{t\geq 0}$ is called CP-divisible iff $V_{t,s}$ is CPTP, and P-divisible iff $V_{t,s}$ is PTP.
\end{Def}

Clearly, any invertible map $\Lambda_t$ is always divisible, since the propagator $V_{t,s}$ can be uniquely defined by $V_{t,s} = \Lambda_t \Lambda_s^{-1}$. Moreover, the propagator $V_{t,s}$ satisfies local composition law

\begin{equation}\label{CL}
  V_{t,u} V_{u,s} = V_{t,s} ,
\end{equation}
for $t \geq u \geq s$. In this case one proves the following

\begin{theorem}[\cite{Angel}] \label{A} Let us assume that $\{\Lambda_t\}_{t\geq0}$ is an invertible dynamical map. Then $\Lambda_t$ is P-divisible iff
\begin{equation}\label{P}
  \frac{d}{dt} \|\Lambda_t(X) \|_1 \leq  0 ,
\end{equation}
for any Hermitian $X \in \mathrm{L}(\mathcal{H})$. It is CP-divisible iff
\begin{equation}\label{CP}
  \frac{d}{dt} \|({\rm id}_d \otimes \Lambda_t)(X)\|_1 \leq  0 ,
\end{equation}
for any Hermitian ${X} \in \mathrm{L}(\mathcal{H} \otimes \mathcal{H})$.
\end{theorem}
Note that if $X$ is traceless then $X = a(\rho_1 - \rho_2)$, where $a \in \mathbb{R}$, and hence (\ref{P}) recovers BLP criterion (\ref{IF}). Interestingly, one proves

\begin{theorem}[\cite{BOGNA}] \label{B} Let us assume that $\{\Lambda_t\}$ is an invertible dynamical map. Then $\{\Lambda_t\}$ is CP-divisible iff
\begin{equation}\label{Bogna}
  \frac{d}{dt} \| [{\rm id}_{d+1} \otimes \Lambda_t]({\rho}_1-{\rho}_2)\|_1 \leq  0 ,
\end{equation}
for any pair of density operators ${\rho}_1$, ${\rho}_2$ in $\mathcal{B}(\mathcal{H}' \otimes \mathcal{H})$ with $\dim(\mathcal{H}')-1=\dim(\mathcal{H})=d$.
\end{theorem}
The essence of (\ref{Bogna}) is that  one enlarges the dimension of the ancilla $d \rightarrow d+1$,  but uses only traceless operators $X = a(\rho_1 - \rho_2)$ like in the original approach to the information flow \cite{BLP}.

For non-invertible maps the divisibility is not guarantied. Actually, one proves

\begin{Proposition}[\cite{PRL-2018}]
  The dynamical  map $\{\Lambda_t\}_{t\geq 0}$ is divisible if and only if

\begin{equation}\label{KK}
  {\rm Ker}(\Lambda_s) \subseteq {\rm Ker}(\Lambda_t) ,
\end{equation}
for any $t > s$.
\end{Proposition}
One has the following generalization of Theorem \ref{A}

\begin{theorem}[\cite{PRL-2018}] \label{B} If the dynamical map $\{\Lambda_t\}_{t\geq 0}$ satisfies
\begin{equation}\label{norm decreasing}
  \frac{d}{dt} \|({\rm id}_d \otimes \Lambda_t)(X)\|_1 \leq  0 ,
\end{equation}
for any Hermitian ${X} \in \mathrm{L}(\mathcal{H} \otimes \mathcal{H})$, then there exists completely positive propagator $V_{t,s} : \LH \to \LH$ which is trace preserving on the image of $\Lambda_s$.
\end{theorem}
Moreover, in the qubit case ($d=2$) one proves

\begin{theorem}[\cite{CC19}] \label{B} The qubit dynamical map $\{\Lambda_t\}_{t\geq 0}$ is CP-divisible iff it satisfies
\begin{equation}\label{}
  \frac{d}{dt} \|({\rm id}_2 \otimes \Lambda_t)(X)\|_1 \leq  0 ,
\end{equation}
for any Hermitian ${X} \in \mathrm{L}(\mathcal{H} \otimes \mathcal{H})$.
\end{theorem}
For $d > 2$ one has the following

{\begin{theorem}[\cite{PRL-2018}] \label{gen div criterion}
	If the dynamical map $\{\Lambda_t\}_{t\geq 0}$ satisfies (\ref{norm decreasing}) as well as the image non-increasing criterion, that is, for any $t\geq s$
	\begin{equation}\label{image non-increasing}
		\rm{Im}{\Lambda_t}\subseteq\rm{Im}{\Lambda_s}
	\end{equation}
	then it is CP-divisible.
\end{theorem}
It should be noted that the theorem above provides a sufficient condition, that might not be necessary (it is not necessary for the qubit case). Whether (\ref{norm decreasing}) is a sufficient condition for CP-divisibility beyond qubit dynamical maps, is still an open question.}

\subsection{Time-discrete maps}

Consider now the family $\{\Lambda_n\}_{n\geq 0}$ of CPTP maps

\begin{equation}\label{}
  \Lambda_n : \mathrm{L}(H_S) \to \mathrm{L}(\mathcal{H}_n)
\end{equation}
where $\mathcal{H}_S,\mathcal{H}_1,\ldots$ are finite dimensional Hilbert spaces of dimensions $d_S,d_1,d_2, \ldots$, respectively. Moreover, we assume that $\Lambda_0 : \mathrm{L}(H_S)\to \mathrm{L}(H_S)$ is an identity map.  We call time discrete dynamical map $\{\Lambda_n\}_{n\geq 0}$ CP-divisible iff there exists a family of CPTP propagators

\begin{equation}\label{Vji}
  V_{j,i} : \mathrm{L}(\mathcal{H}_i) \to \mathrm{L}(\mathcal{H}_j) , \ \ \ (j > i)
\end{equation}
such that
\begin{equation}\label{}
  \Lambda_j = V_{j,i} \Lambda_i \ .
\end{equation}
Consider now an ensemble of states $\rho^x$ prepared with probability $p_x$: $\mathcal{E} =\{p_x,\rho^x\}_x$.
To distinguish between these states one defines guessing probability

\begin{equation}\label{}
  \mathrm{P}_{\rm guess}(\mathcal{E}) = \max \sum_{x \in \mathcal{X}} p_x {\rm Tr}(P^x \rho^x) ,
\end{equation}
where the maximum is over all POVMs $\{P^x\}_x$ defined on the Hilbert space $\mathcal{H}$.  Buscemi and Datta  \cite{datta} introduced the following interesting concept enabling one to compare quantum channels

\begin{Def} Time discrete dynamical map $\{\Lambda_n\}_{n\geq 0}$ is information decreasing iff for any $j > i $ one has
  \begin{equation}\label{PjPi}
     \mathrm{P}_{\rm guess}(\mathcal{E}_j) \leq  \mathrm{P}_{\rm guess}(\mathcal{E}_i) ,
  \end{equation}
where $\mathcal{E}_n =\{p_x,\Lambda_n(\rho^x)\}_x$. $\{\Lambda_n\}_{n\geq 0}$ is completely information decreasing iff ${\rm id}_S \otimes \Lambda_n$ is information decreasing (${\rm id}_S$ stands for the identity map on $\mathrm{L}(\mathcal{H}_S)$).
\end{Def}

\begin{theorem}[\cite{datta}] Time discrete dynamical map $\{\Lambda_n\}_{n\geq 0}$ is CP-divisible iff it is completely information decreasing.
\end{theorem}
Note, that if $\mathcal{E}$ consists of two members $\mathcal{E}=\{p_1,\rho_1;p_2,\rho_2\}$, then by \cite{HEL} (see also \cite{JUNU} for the review)

\begin{equation}\label{}
     \mathrm{P}_{\rm guess}(\mathcal{E}) = \frac 12 \| p_1 \rho_1 - p_2 \rho_2 \|_1 ,
\end{equation}
and hence the monotonicity condition (\ref{PjPi})

\begin{equation}\label{ppPP}
      \| [{\rm id}_S \otimes \Lambda_j](X) \|_1 \leq   \|  [{\rm id}_S \otimes \Lambda_i](X) \|_1 ,
\end{equation}
{for any hermitian $X$ in $\textrm{L}(\mathcal{H}_S\otimes\mathcal{H}_S)$}
defines a necessary condition for CP-divisibility.

\begin{Proposition} If ${\rm Im}(\Lambda_n) = \mathrm{L}(\mathcal{H}_n)$ for all $n > 0$, then $\{ \Lambda_n\}_{n\geq 0}$ is CP-divisible iff (\ref{ppPP}) is satisfied for all $j > i$.
\end{Proposition}
Proof: We show that (\ref{ppPP}) implies CP-divisibility. Note, that  $\{ \Lambda_n\}_{n\geq 0}$ is divisible, that is, there exists  the family $V_{j,i}$ satisfying (\ref{Vji}). 
We can define

\begin{equation}\label{LLR}
  V_{j,i} := \Lambda_j \Lambda_i^{-{1_R}},
\end{equation}
where $\Lambda_i^{-1_{R}}$ denotes a (linear) right inverse of $\Lambda_i$, that is, one has $\Lambda_i \Lambda_i^{-{1_R}}= \oper_{\mathcal{H}_i}$, but in general $\Lambda_i^{-{1_R}} \Lambda_i\neq \oper_{\mathcal{H}_S}$. A right inverse exists, because ${\rm Im}(\Lambda_i) = \mathrm{L}(\mathcal{H}_i)$, {that is, $\Lambda_i$ is surjective}. A left inverse does not exist unless $\rm{Ker}(\Lambda_i)=\{0\}$, and we do not require it to exist here either.

We see that $V_{j,i}$ defined thus is a valid propagator: For all $X\in \rm{L}(\mathcal{H}_S)$
\begin{equation}
\Lambda_{i}^{-1_{R}}\Lambda_{i}(X)=X+\delta
\end{equation}
for some $\delta\in\rm{Ker}(\Lambda_{i})$, and hence

\begin{equation}
V_{j,i}\Lambda_{i}(X)=(\Lambda_{j}\Lambda_{i}^{-1_{R}})\Lambda_{i}(X)=\Lambda_j(X+\delta)=\Lambda_j(X)
\end{equation}
as $\rm{Ker}(\Lambda_i) \subseteq \rm{Ker}(\Lambda_j)$

Now, for any Hermitian $X \in \mathrm{L}(\mathcal{H}_S \otimes \mathcal{H}_S)$ one  has

\begin{equation}\label{}
  \| [{\rm id}_S \otimes \Lambda_j](X) \|_1 = \| [{\rm id}_S \otimes V_{j,i} \Lambda_i](X) \|_1 \leq \| [{\rm id}_S \otimes\Lambda_i](X)\|_1 ,
\end{equation}
which means that

\begin{equation}\label{}
   \| [{\rm id}_S \otimes V_{j,i}](Y) \|_1 \leq \|Y\|_1 ,
\end{equation}
for any $Y = [{\rm id}_S \otimes\Lambda_i](X)$. Now, since ${\rm Im}(\Lambda_i) = \mathrm{L}(\mathcal{H}_i)$, $V_{j,i}$ is trace-preserving and ${\rm id}_S \otimes V_{j,i}$ defines a contraction w.r.t. trace norm, and thus {by Lemma 1 of \cite{PRL-2018},} one finds that $V_{j,i}$ is CPTP (note that ${\rm dim}\, \mathcal{H}_S \geq {\rm dim}\, \mathcal{H}_n$ for $n > 0$). \hfill $\Box$

Clearly, $\Lambda_i^{-{1_R}}$ is not uniquely defined but it is trace-preserving. Note however that

\begin{Proposition} The corresponding propagator $V_{j,i}$ (\ref{LLR}) is uniquely defined, independent of the choice of right inverse $\Lambda_i^{-1_R}$.
\end{Proposition}
Proof: Suppose that there are two propagators $V_{j,i}$ and $\tilde{V}_{j,i}$ defined via

$$  V_{j,i} := \Lambda_j \Lambda_i^{-{1_R}} \ , \ \ \ \tilde{V}_{j,i} := \Lambda_j \tilde{\Lambda_i}^{-{1_R}} , $$
where $\Lambda_i^{-{1_R}} $ and $\tilde{\Lambda_i}^{-{1_R}}$ are two different right inverses of $\Lambda_i$. Now, let

$$   X =  \Lambda_i^{-{1_R}}(Y) \ , \ \ \  \tilde{X} = \tilde{\Lambda_i}^{-{1_R}}(Y) . $$
Note, that $\Lambda_i \Lambda_i^{-{1_R}}(Y) = Y$ and $\Lambda_i   \tilde{\Lambda_i}^{-{1_R}}(Y) = Y$, and hence

$$   \Lambda_i(X) = Y= \Lambda_i(\tilde{X}) , $$
which shows that $X- \tilde{X} \in {\rm Ker}(\Lambda_i)$. Finally, divisibility requires that $ {\rm Ker}(\Lambda_j) \supseteq {\rm Ker}(\Lambda_i)$ and hence $\Lambda_j(X) = \Lambda_j(\tilde{X})$ which implies  $V_{j,i}(Y) = \tilde{V}_{j,i}(Y)$.  \hfill $\Box$

\vspace{.1cm}

\begin{Cor} In particular, if $\mathcal{H}_S = \mathcal{H}_n$ for $n>0$, and $\Lambda_n$ is invertible, then the map  $\{ \Lambda_n\}_{n\geq 0}$ is CP-divisible iff (\ref{ppPP}) holds for all Hermitian $X \in \mathrm{L}(\mathcal{H}_S \otimes \mathcal{H}_S)$.
\end{Cor}

\section{Generalized inverse}

For invertible maps the propagator $V_{t,s}$ is uniquely defined via $V_{t,s}=\Lambda_t \Lambda_s^{-1}$. Suppose now that the dynamical map $\{\Lambda_t\}_{t\geq 0}$ is divisible but not necessarily invertible. There is a natural construction of a propagator via the following formula

\begin{equation}\label{PRO}
  V_{t,s} = \Lambda_t \Lambda_s^- ,
\end{equation}
where $\Lambda_s^- : \LH \to \LH$ denotes a generalized inverse of $\Lambda_s$. Let us briefly recall the concept of a generalized inverse (cf. \cite{Kumar,GI,Horn,Yanai}). Consider a linear operator $A : \mathcal{H}_1 \to \mathcal{H}_2$ acting between two Hilbert spaces.
One defines a generalized inverse \cite{GI,Horn} of $A$ to be a linear operator $A^- : \mathcal{H}_2 \to \mathcal{H}_1$ satisfying the defining property

\begin{equation}\label{}
  AA^-A = A .
\end{equation}
Clearly, $A^-$ is arbitrarily defined outside the image of A. A generalized inverse $A^-$ is called reflexive if additionally

\begin{equation}\label{R}
  A^-AA^- = A^- .
\end{equation}
Actually, the additional property (\ref{R}) is  equivalent to ${\rm Rank}(A^-)= {\rm Rank}(A)$ (in general one has ${\rm Rank}(A^-) \geq {\rm Rank}(A)$). It turns out \cite{GI} that given any generalised inverse $A^-$ of $A$, there exist transversal subspaces $V$ and $W$ of $\rm{Ker}(A)$ and $\rm{Im}(A)$ respectively, that is,

$$ \mathcal{H}_1 = {\rm Ker}(A) \oplus V \ , \ \ \ \mathcal{H}_2 = W \oplus {\rm Im}(A) , $$
with the induced decomposition

$$ x = x_0 \oplus x_1 \in \mathcal{H}_1 \ , \ \ \ y = y_0 \oplus y_1 \in \mathcal{H}_2 , $$
and a linear operator  $B:W \to \rm{Ker}(A)$, such that if $Ax = y$, then

\begin{equation}\label{}
A^- y = B y_0 + x_1 .
\end{equation}
One finds therefore that

\begin{equation}\label{}
AA^- = \Pi_{{\rm Im}(A)} ,
\end{equation}
defines a projector onto ${\rm Im}(A)$ along $W$, and

\begin{equation}\label{}
A^-A = \Pi_{V} ,
\end{equation}
defines a projector onto $V$ along $\rm{Ker(A)}$. Different choices of transversal subspaces $V$ and $W$ provide different generalized inverses. Now, there are two special choices of subspaces $V$ and $W$, that is, when $V \perp {\rm Ker}(A)$ and $W \perp {\rm Im}(A)$. Let $A^\dagger : \mathcal{H}_2 \to \mathcal{H}_1$ denote the adjoint of $A$. One has

\begin{Proposition} The following conditions are equivalent
	
	\begin{itemize}
		\item $V \perp {\rm Ker}(A)$,
		
\item  $V= {\rm Im}(A^\dagger )$,

\item $A^-A$ is an orthogonal projector, that is,

\begin{equation}\label{I}
	(A^-A)^\dagger = A^-A .
\end{equation}
\end{itemize}
Similarly, the following conditions are equivalent

\begin{itemize}
  \item $W \perp {\rm Im}(A)$
  \item $W= {\rm Ker}(A^\dagger )$
  \item $AA^-$ is an orthogonal projector, that is,

\begin{equation}\label{II}
	(AA^-)^\dagger = AA^- .
\end{equation}
\end{itemize}
\end{Proposition}
Actually, there is a unique generalized inverse satisfying (\ref{R}), (\ref{I}) and (\ref{II}) and it is called Moore-Penrose generalized inverse. Moore-Penrose inverse is uniquely defined and clearly reduces to the standard inverse when $A$ is invertible.

Consider now a singular value decomposition (SVD) of the complex $d_2 \times d_1$ matrix $A$

\begin{equation}\label{}
  A = U \Sigma V^\dagger ,
\end{equation}
where $U \in U(d_2)$ and $V\in U(d_1)$ (where $U(d)$ denotes the set of all unitary $d \times d$ matrices), and $\Sigma$ is a $d_2 \times d_1$ matrix displaying the following block structure

\begin{equation}\label{}
  \Sigma = \left( \begin{array}{cc}
                    D & 0 \\
                    0 & 0
                  \end{array} \right) ,
\end{equation}
and $D$ is a diagonal $r\times r$ matrix with positive (diagonal) entries ($r \leq \min\{d_1,d_2\}$). Any generalized inverse of $A$ has the following form

\begin{equation}\label{}
  A^- = V \Sigma^- U^\dagger ,
\end{equation}
with

\begin{equation}\label{S}
  \Sigma^- = \left( \begin{array}{cc}
                    D^{-1} & X \\
                    Y & Z
                  \end{array} \right) ,
\end{equation}
and $X,Y,Z$ are completely arbitrary matrices with appropriate dimensions.

\begin{Proposition} Generalized inverse $A^-$

\begin{enumerate}
  \item satisfies (\ref{R}) iff $Z=YDX$,
  \item satisfies (\ref{I}) iff $Y=0$,
  \item satisfies (\ref{II}) iff $X=0$,
  \item is Moore-Penrose generalized inverse iff $X=0$, $Y=0$ and $Z=0$.
\end{enumerate}

\end{Proposition}

\section{Construction of propagators}

Equipped with the notion of generalized inverse we are going to analyze propagators defined via

\begin{equation}\label{}
  V_{t,s} = \Lambda_t \Lambda_s^- ,
\end{equation}
with $\Lambda_s^- : \LH \to \LH$ being a generalized inverse of $\Lambda_s$, that is,

\begin{equation}\label{LLL=L}
  \Lambda_s \Lambda_s^- \Lambda_s = \Lambda_s .
\end{equation}
Clearly, $\Lambda_s^-$ is not uniquely defined.  Suppose, that the image of $\Lambda_s$ is a proper subspace of $\LH$, and let $\mathcal{C}_s$ be a complementary subspace such that for any $s$ one has

\begin{equation}\label{C}
  \LH =   \mathcal{C}_s \oplus {\rm Im}(\Lambda_s) .
\end{equation}
It should be stressed that {in general} $\mathcal{C}_s$ is not uniquely defined. Actually, any linear subspace $\mathcal{C}_s$ such that ${\rm dim}(\mathcal{C}_s) + {\rm dim} ({\rm Im}(\Lambda_s)) = d^2$, and $\mathcal{C}_s \cap {\rm Im}(\Lambda_s) = {\{0\}}$ does the job {of giving us a propagator, and in general different propagators can be expected to correspond to different $\mathcal{C}_s$}. Now, {for any $Y \in \LH$ we have the unique decomposition}

$$  Y = Y_0 \oplus Y_1 \in  \mathcal{C}_s \oplus {\rm Im}(\Lambda_s)  , $$
{where, by the last statement, we mean that $Y_1 \in {\rm Im}(\Lambda_s)$ and $Y_0 \in \mathcal{C}_s$.} One has

\begin{equation}\label{}
  \Lambda_s^-(Y) := X_1 + \Phi_s(Y_0) ,
\end{equation}
where {$\Lambda_s(X_1)=Y_1$} (that is, $X_1$ comes from decomposing $\textrm{L}(\mathcal{H})$ into ${\rm Ker}(\Lambda_s)$ and a transversal subspace), and $\Phi_s : \mathcal{C}_s \to {\rm Ker}(\Lambda_s)$  is an arbitrary linear map. The arbitrary map $\Phi_s$ defines a gauge freedom of the whole construction, i.e. $\Lambda_s$ is defined up to the arbitrary gauge function $\Phi_s$. Now, one has for the propagator

\begin{equation}\label{37}
  V_{t,s}(Y) = \Lambda_t\Lambda_s^-(Y) = \Lambda_t(X_1 + \Phi_s(Y_0)) = \Lambda_t(X_1) ,
\end{equation}
due to $\Phi_s(Y_0) \in {\rm Ker}(\Lambda_s) \subseteq {\rm Ker}(\Lambda_t)$. Hence {such a} propagator is {characterized} by the family of subspaces $\{\mathcal{C}_t\}_{t \geq 0}$ transversal to images of $\{\Lambda_t\}_{t \geq 0}$. All such propagators agree on $\rm{Im}(\Lambda_s)$, due to the kernel non-decreasing property (\ref{KK}). Interestingly, $\Lambda^-_s$ depends upon gauge function. but the propagator is perfectly gauge-invariant, {and may only depend on the choice of $\mathcal{C}_s$}. Note, that formula (\ref{37}) shows that

\begin{equation}\label{}
  V_{t,s}\Lambda_s(X) = \Lambda_t(X) ,
\end{equation}
which is a defining property of the propagator.

{\begin{Proposition}\label{PRO-CL}  A propagator $V_{t,s}=\Lambda_t \Lambda_s^-$ satisfies composition law (\ref{CL}).
\end{Proposition}}
Proof: Indeed, one has

\begin{equation}
V_{t,u}V_{u,s} = V_{t,u}(\Lambda_{u}\Lambda_{s}^-) = (V_{t,u} \Lambda_{u}) \Lambda_{s}^- = \Lambda_{t}\Lambda_{s} ^- = V_{t,s} ,
\end{equation}
which ends the proof. \hfill $\Box$

{\begin{Proposition}
  $V_{t,s} = \Lambda_t \Lambda_s^-$ is  trace-preserving  iff the subspace $\mathcal{C}_s$ contains only traceless operators.
\end{Proposition}}
Proof: One finds

\begin{equation*}\label{}
  {\rm Tr}\,  V_{t,s}(Y) = {\rm Tr}\, \Lambda_t(X_1) = {\rm Tr}\, \Lambda_s(X_1)  ,
\end{equation*}
due to trace-preservation of the maps $\Lambda_t$ and $\Lambda_s$. Now, using $Y = \Lambda_s(X_1) \oplus Y_0$, one {f}inds $ {\rm Tr}\, \Lambda_s(X_1) = {\rm Tr}\, Y - {\rm Tr}\, Y_0  = {\rm Tr}\, Y$,
if and only if ${\rm Tr}\,Y_0=0$.   \hfill $\Box$


Note, that $V_{t,s}= \Lambda_t \Lambda_s^-$ satisfies

\begin{equation}\label{image}
  {\rm Im}(V_{t,s}) =  {\rm Im}(\Lambda_t) \subseteq \LH ,
\end{equation}
for any $t\geq s$.

\begin{Proposition}
	If the propagator $V_{t,s}$ satisfy  (\ref{image}), then $V_{t,s}=\Lambda_t \Lambda_s^-$ for some generalized inverse $\Lambda_s^-$.
\end{Proposition}
Proof: Note,  that the most general propagator satisfying (\ref{image}) reads as follows

\begin{equation}\label{}
  V_{t,s}(Y) = \Lambda_t(X_1) + \Phi_{t,s}(Y_0),
\end{equation}
where $X_1$ satisfies $Y_1 = \Lambda_s(X_1) \in {\rm Im}(\Lambda_t)$, and  $\Phi_{t,s} : \mathcal{C}_s \to {\rm Im}(\Lambda_t)$
is {some} linear map. One has therefore $V_{t,s}(Y)=\Lambda_t(X_1 + \delta)$, where $\Lambda_t(\delta) = \Phi_{t,s}(Y_0) \in {\rm Im}(\Lambda_t)$. Hence,  $V_{t,s}(Y)=\Lambda_t(X_1')$, where $X_1'=X_1 + \delta$ corresponds to another decomposition $Y=Y_1' \oplus Y_0'$, w.r.t. the new family of transversal subspaces $\{\mathcal{C}'_t\}_{t \geq 0}$, {with $Y_1' = \Lambda_s(X_1')$} \hfill $\Box$

\vspace{.1cm}

{As we see in the proof above, the key idea is to choose a ``good" family of transversal subspaces to decompose into. The proposition above essentially says that such a decomposition always exists if the condition (\ref{image}) is satisfied.} Clearly, the same techniques applies for the time-discrete dynamical maps $\{\Lambda_n\}_{n\geq 0}$. One defines

\begin{equation}\label{}
  V_{j,i} := \Lambda_j \Lambda_i^- ,
\end{equation}
which reduces to $ \Lambda_j \Lambda_i^{-1_{R}}$ iff ${\rm Im}(\Lambda_i) = \mathrm{L}(\mathcal{H}_i)$.

Finally, one has the following
{\begin{Proposition}
	For CP-divisible dynamical maps satisfying (\ref{image non-increasing}), there exists a CPTP propagator $V_{t,s} = \Lambda_t \Lambda_s^{-}$ for some  generalized inverse$\Lambda_s^{-}$.
\end{Proposition}
Proof: For a CP-divisible dynamical map $\{\Lambda_t\}_{t\geq 0}$ that satisfies (\ref{image non-increasing}), given that the rank of the map decreases at times $t_1 < t_2 < ... < t_n$, there exist CPTP projections $\Pi_{t_i} : \rm{Im}{\Lambda_{t_{i-1}}} \to \rm{Im} {\Lambda_{t_{i}}}$, (with $i=1,2,...,n$ and $t_0=0$) by \cite{PRL-2018}. Hence, if $V_{t,s}:\rm{L}(\mathcal{H})\to \rm{L}(\mathcal{H})$ is a CPTP propagator, then, for $m$ such that $t_{m+1} > t \geq t_{m}$, $\Pi_{t_{m}} \Pi_{t_{m-1}} ... \Pi_{t_{1}} V_{t,s}$ is a CPTP propagator that satisfies (\ref{image}), and hence admits a construction through a generalised inverse.
\hfill $\Box$}


\vspace{.1cm}







\section{Super-operators -- matrix representation of propagators}

Using well known vectorisation procedure \cite{Watrous, Gilchrist} which assigns to any operator $X \in \LH$ a vector $|X\>\!\> \in \mathcal{H} \otimes \mathcal{H}$

\begin{equation}\label{}
  X \to |X\>\!\> =\sum_{i,j} X_{ij} |i \otimes j\>
\end{equation}
with $X_{ij} = \<i|X|j\>$, one may assign to any linear map $\Phi : \LH \to \LH$ a super-operator $\widehat{\Phi} : \mathcal{H} \otimes \mathcal{H} \to \mathcal{H} \otimes \mathcal{H}$ defined by

\begin{equation}\label{}
  \widehat{\Phi}|X\>\!\> := |\Phi(X)\>\!\> .
\end{equation}
One easily finds \cite{Watrous} that

{\begin{equation}\label{}
  |A X B\>\!\> = A \otimes B^T |X\>\!\> ,
\end{equation}
and
\begin{equation}\label{}
(A,B)=\<\!\<A|B\>\!\>
\end{equation} for any $A,B \in \LH$, $(A,B) := {\rm Tr}(A^\dagger B)$ being the Hilbert-Schmidt inner product}. This assignment enjoys a fundamental property

\begin{equation}\label{}
  \widehat{\Phi_1 \Phi}_2 = \widehat{\Phi}_1 \widehat{\Phi}_2 ,
\end{equation}
and hence
\begin{equation}\label{}
  \widehat{\Phi^{-1}} = \widehat{\Phi}^{-1} .
\end{equation}
Note, that
\begin{equation}\label{}
  \widehat{\Phi^\ddag} = \widehat{\Phi}^\dagger
\end{equation}
where $\Phi^\ddag$ is the adjoint (Heisenberg picture) of $\Phi$ defined via
\begin{equation}\label{}
  (\Phi(X),Y) = (X,\Phi^\ddag(Y)) ,
\end{equation}
for all $X,Y \in \LH$.
{Note that
\begin{equation}
\<\!\<\oper|X\>\!\>={\rm Tr}\,X
\end{equation}
Thus,} the map $\Phi$ is trace-preserving if and only if

\begin{equation}\label{}
  \widehat{\Phi}^\dagger  |\oper\>\!\> = |\oper\>\!\> ,
\end{equation}
where $ |\oper\>\!\> = \sum_i |i \otimes i\>$. It is unital, i.e. $\Phi(\oper) = \oper$, if and only if
\begin{equation}\label{}
  \widehat{\Phi}  |\oper\>\!\> = |\oper\>\!\> .
\end{equation}
The propagator may be, therefore, equivalently represented via

\begin{equation}\label{}
  \widehat{V}_{t,s} = \widehat{\Lambda}_t \widehat{\Lambda}^-_s .
\end{equation}

\begin{Example}   \label{EX1}

Consider the following dynamical map

\begin{equation}\label{}
  \Lambda_t(\rho) = [1-f(t)] \rho + f(t) \omega_t {\rm Tr}\, \rho
\end{equation}
where $\omega_t$ is a time-dependent density operator, and $f : \mathbb{R}_+ \to [0,1]$ with the initial condition $f(0)=0$. Now, the map is invertible iff $f(t) < 1$  for all $t$. Suppose, that $f(t_*)=1$ for some $t_*>0$. Then the map is divisible iff $f(t)=1$ for all $t\geq t_*$. Hence, for $t\geq t_*$ the map $\Lambda_t(\rho) = \omega_t {\rm Tr}\, \rho$ projects any density operator into $\omega_t$ and clearly is not invertible. Equivalently, the map can be represented via the corresponding family of super-operators

\begin{equation}\label{}
  \widehat{\Lambda}_t = [1-f(t)] \oper \otimes \oper + f(t) |\omega_t\> \!\> \<\!\< \oper | .
\end{equation}
Now, to define the propagator $ \widehat{V}_{t,s} = \widehat{\Lambda}_t \widehat{\Lambda}^-_s$ for $s \geq t_*$ one needs to define the generalized inverse $\widehat{\Lambda}^-_s$ which is highly non unique and depends upon the family of subspaces $\mathcal{C}_s$ transversal to the image of $\Lambda_s$.

 There are two natural choices of $\mathcal{C}_s$. Taking $\mathcal{C}_t = {\rm Ker}(\Lambda_t) = \{ X \, |\, {\rm Tr}X=0\}$, or equivalently $\<\!\< \oper|X\>\!\>=0$, one finds (assuming that the gauge function vanishes)
\begin{equation}\label{a}
  \widehat{\Lambda}^-_s = |\omega_s\>\!\> \<\!\< \oper| ,
\end{equation}

and hence

\begin{equation}\label{Va}
  \widehat{V}_{t,s} = |\omega_t\>\!\> \<\!\< \oper|.
\end{equation}
The second natural choice is ${\rm Ker}(\Lambda^\ddag_t) = \{ X \, |\, {\rm Tr}(\omega_t X)=0\}$, or equivalently $\<\!\< \omega_s|X\>\!\>=0$, one finds (again with vanishing gauge function) generalized inverse to be

\begin{equation}\label{b}
  \widehat{\Lambda}^-_s = \frac{ |\omega_s\>\!\> \<\!\< \omega_s| }{ \<\!\< \omega_s| \omega_s\>\!\> }   ,
\end{equation}
which implies

\begin{equation}\label{Vb}
  \widehat{V}_{t,s} = \frac{ |\omega_t\>\!\> \<\!\< \omega_s| }{ \<\!\< \omega_s| \omega_s\>\!\> } .
\end{equation}
Interestingly, computing the Moore-Penrose generalized inverse one finds

{
\begin{equation}\label{c}
  \widehat{\Lambda}^{\rm MP}_s = \frac{ |\oper \>\!\> \<\!\< \omega_s| }{ \<\!\< \omega_s| \omega_s\>\!\> \<\!\< \oper| \oper\>\!\> }   ,
\end{equation}
}
which again implies propagator (\ref{Vb}). The properties of the above three generalized inverses (\ref{a}), (\ref{b}), and (\ref{c}) are summarized in the table.

\begin{table}[]\label{table1}
\begin{tabular}{|l|l|l|l|l|}
\hline {Gen. inv.}
 & reflexive &  (\ref{I}) &  (\ref{II}) & {TP} \\  \hline
(\ref{a}) & yes  & no & no & yes \\ \hline
(\ref{b}) & yes  & no & yes & no \\ \hline
(\ref{c}) & yes & yes & yes & no \\ \hline
\end{tabular}
\end{table}

{\begin{table}[]\label{table2}
	\begin{tabular}{|l|l|l|}
		\hline Propagator
		&  CP &  TP \\
		\hline
		(\ref{Va}) & yes  & yes \\ \hline
		(\ref{Vb}) & yes  & no  \\ \hline
	\end{tabular}
\end{table}    }
Note, that the propagator (\ref{Va}) is CPTP but (\ref{Vb}) is CP but not trace-preserving.

\end{Example}






\section{Propagator from spectral properties of dynamical maps}

Generalized inverse $\Lambda_s^-$ is highly non unique. There is, however, a natural way to define a generalized inverse (and hence the propagator) using spectral properties of the dynamical map. In this section we analyze both diagonalizable and non-diagonalizable case as well.

\subsection{Diagonalizable dynamical maps}

Let us assume that dynamical map $\{\Lambda_t\}$ is diagonalizable, that is, for any $t \geq 0$ one has

\begin{equation}\label{LL*}
  \widehat{{\Lambda}_t} |F_\alpha(t)\>\!\> = \lambda_\alpha(t) |F_\alpha(t)\>\!\>  \ ,
\end{equation}
and
\begin{equation}
  \widehat{\Lambda_t}^\dagger |G_\alpha(t)\>\!\> = \lambda^*_\alpha(t) |G_\alpha(t)\>\!\>  \ ,
\end{equation}
{for some $F_\alpha (t), G_\alpha (t) \in L(H)$},  $\alpha=0,1,\ldots,d^2-1$.  The corresponding right and left eigenvectors satisfy

\begin{equation}\label{}
  \<\!\< G_\alpha(t)|F_\beta(t) \>\!\>   = \delta_{\alpha\beta} .
\end{equation}
One has for the spectral resolution

\begin{equation}\label{}
\widehat{\Lambda_t} = \sum_{\alpha=0}^{d^2-1}\lambda_\alpha(t) \widehat{\mathcal{P}}_\alpha(t) ,
\end{equation}
where
\begin{equation}\label{}
  \widehat{\mathcal{P}}_\alpha(t)(\rho) :=  | F_\alpha(t)  \>\!\> \<\!\<   G_\alpha(t)| ,
\end{equation}
are projectors (not necessarily Hermitian) satisfying

\begin{equation}\label{PPP}
  \widehat{\mathcal{P}}_\alpha(t) \widehat{\mathcal{P}}_\beta(t) = \delta_{\alpha\beta}  \widehat{\mathcal{P}}_\alpha(t) .
\end{equation}
Note, that trace-preservation condition implies  $\lambda_0(t)=1$ and $G_0(t) = \oper$. Suppose now that the eigenvalues satisfy $\lambda_\alpha(s)=0$ for $\alpha > n$ {and are non-zero for $\alpha \leq n$}. Divisibility implies that $\lambda_\alpha(t)=0$ for $\alpha > n$ for all $t > s$. Let us define generalised inverse via

\begin{equation}\label{inv-spec}
  \widehat{{\Lambda}}^-_s = \sum_{\alpha=0}^{n}\lambda^{-1}_\alpha(s) \widehat{\mathcal{P}}_\alpha(s)  .
\end{equation}

The corresponding propagator reads

\begin{equation}\label{}
\widehat{V}_{t,s} = \sum_{\alpha,\beta=0}^{n} \frac{\lambda_\alpha(t)}{\lambda_\beta(s)} \,  C_{\alpha\beta}(t,s)\, |F_\alpha(t)  \>\!\> \<\!\< G_\beta(s)|  ,
\end{equation}
where the `correlator' $C_{\alpha\beta}(t,s)$ reads

\begin{equation}\label{}
C_{\alpha\beta}(t,s) =     \<\!\<  G_\alpha(t)| F_\beta(s) \>\!\> .
\end{equation}


\begin{Proposition} Generalized inverse defined via (\ref{inv-spec}) is reflexive, that is, $\Lambda_t^- \Lambda_t \Lambda_t^- = \Lambda_t^-$.
Moreover, both projectors

$$ \widehat{{\Lambda}}_t \widehat{{\Lambda}}_t^- = \widehat{{\Lambda}}_t^- \widehat{{\Lambda}}_t  = \sum_{\alpha=0}^{n}   \widehat{\mathcal{P}}_\alpha(t) \,   $$
project to the image of ${\Lambda}_t$.
\end{Proposition}

\begin{Cor} If $\Lambda_t$ is normal, that is, $F_\alpha(t) = G_\alpha(t)$, then generalized inverse defined via (\ref{inv-spec}) is the Moore-Penrose generalized inverse.
\end{Cor}

The corresponding propagator $\Lambda_t \Lambda_s^-$ is evidently trace-preserving. Moreover, in this case the family of subspaces $\mathcal{C}_t$ satisfy $\mathcal{C}_t = {\rm Ker}(\Lambda_t)$.

\begin{Example}  \label{EX3}
Let us observe that dynamical map in {Example 1} is diagonalizable. Indeed, one finds

\begin{equation}\label{}
  \widehat{\Lambda}_t|\omega_t \>\!\>  = |\omega_t \>\!\>  , \ \ \ \widehat{\Lambda}_t |F_\alpha\>\!\>  = [1-f(t)] |F_\alpha \>\!\>,
\end{equation}
where $F_\alpha$ is traceless for $\alpha=1,\ldots,d^2-1$. The dual map reads

\begin{equation}\label{}
  \widehat{\Lambda}_t^\dagger = [1-f(t)] \oper \otimes \oper   + f(t) \, | \oper \>\!\>  \<\!\<  \omega_t|  ,
\end{equation}
and hence

\begin{equation}\label{}
  \widehat{\Lambda}^\dagger_t |\oper \>\!\>  = |\oper \>\!\>  , \ \ \ \widehat{\Lambda}_t |G_\alpha\>\!\>  = [1-f(t)] |G_\alpha\>\!\>  ,
\end{equation}
where $G_\alpha$ satisfy $\<\!\< G_\alpha |\omega_t\>\!\> =0$. Hence the spectral resolution reads as follows

\begin{equation}\label{}
  \widehat{\Lambda}_t  = |\omega_t\>\!\> \<\!\< \oper| + [1-f(t)] \sum_{\alpha=1}^{d^2-1}  | F_\alpha\>\!\> \<\!\< G_\alpha| .
\end{equation}
Now, for $s \geq t_*$, it reduces to

\begin{equation}\label{}
  \widehat{\Lambda}_s =  |\omega_s\>\!\> \<\!\< \oper|
\end{equation}
that is, there is only one non-vanishing eigenvalue $\lambda_0(t)=1$. Hence, the formula (\ref{inv-spec}) leads to

\begin{equation}\label{}
  \widehat{\Lambda}^-_s(\rho)  =  |\omega_s\>\!\> \<\!\< \oper| ,
\end{equation}
which reproduces (\ref{a}).

\end{Example}

\subsection{Non-diagonalizable dynamical maps}

Consider now a general case corresponding to the following spectral Jordan decomposition \cite{Kato}

\begin{equation}\label{}
\widehat{{\Lambda}}_t = \sum_{\alpha=0}^{d^2-1} (\lambda_\alpha(t) \widehat{\mathcal{P}}_\alpha(t) + \widehat{\mathcal{N}}_\alpha(t) ),
\end{equation}
where $\mathcal{N}_\alpha(t)$ are nilpotent maps satisfying

\begin{equation}\label{NPN}
  \widehat{\mathcal{P}}_\alpha(t) \widehat{\mathcal{N}}_\beta(t) =  \widehat{\mathcal{N}}_\beta(t) \widehat{\mathcal{P}}_\alpha(t) = \delta_{\alpha\beta}  \widehat{\mathcal{N}}_\alpha(t) ,
\end{equation}
and $ \widehat{\mathcal{P}}^{n_\alpha}_\alpha(t) =0$,  with $1\leq n_\alpha \leq {\rm rank}\,  \widehat{\mathcal{P}}_\alpha(t)$.  Suppose that for $t \geq t_*$ the eigenvalues satisfy $\lambda_\alpha(t)=0$ for $\alpha > n$. One has therefore

\begin{equation}\label{}
\widehat{{\Lambda}}_t = \sum_{\alpha=0}^{n} (\lambda_\alpha(t) \widehat{\mathcal{P}}_\alpha(t) + \widehat{\mathcal{N}}_\alpha(t) ) + \sum_{\alpha=n+1}^{d^2-1} \widehat{\mathcal{N}}_\alpha(t)  .
\end{equation}
Clearly nilpotent maps are represented by Jordan blocks.

\begin{Proposition} \label{PRO-JORDAN} Let $J_k(\lambda)$ be a Jordan block of size $k$. One finds
\begin{equation}\label{Jordan}
  J_k(0) J^T_k(0) J_k(0) = J_k(0) ,
\end{equation}
and
\begin{equation}\label{Jordan2}
  J^T_k(0) J_k(0) J^T_k(0) = J^T_k(0) ,
\end{equation}
and hence $J^T_k(0)$ is a reflexive generalized inverse of $J_k(0)$.
\end{Proposition}

%

\section{Bloch representation and Propagators}

For a qubit system one often uses well known Bloch representation

\begin{equation}\label{}
  \rho = \frac{1}{2} (\oper + \sum_{k=1}^3 r_k \sigma_k) ,
\end{equation}
where {$\mathbf{r}=(r_1,r_2,r_3)^T$} is the Bloch vector corresponding to $\rho$. Now, for a qubit map $\Phi$ one defines a real $4 \times 4 $ matrix

\begin{equation}\label{}
\Phi_{\alpha\beta} := \frac 12 {\rm Tr}(\sigma_\alpha \Phi(\sigma_\beta)) ,
\end{equation}
with $\sigma_0=\oper$. The matrix $\Phi_{\alpha\beta}$ has the following structure

\begin{equation}\label{}
  \Phi_{\alpha\beta} = \left( \begin{array}{c|c} 1 & 0 \\ \hline  \mathbf{x} & \Delta \end{array} \right) ,
\end{equation}
{where $\mathbf{x} \in \mathbb{R}^3$, $\Delta$ is a $3\times 3$ real matrix}, and the map $\rho \to \Phi(\rho)$ in the Bloch representation is realized via the following affine transformation

\begin{equation}\label{}
  \mathbf{r} \to {\Delta} \mathbf{r} + \mathbf{x} .
\end{equation}
This representation may be generalized for arbitrary dimension $d$: let $\tau_\alpha$ ($\alpha=0,1,\ldots,d^2-1$) be Hermitian orthonormal basis in $\LH$ such that $\tau_0 = \oper/\sqrt{d}$. Any CPTP map $\Phi : \LH \to \LH$ gives rise to real $d^2\times d^2$ matrix

\begin{equation}\label{}
\Phi_{\alpha\beta} =  {\rm Tr}(\tau_\alpha \Phi(\tau_\beta)) .
\end{equation}
Consider now the dynamical map $\{\Lambda_t\}_{t \geq 0}$. One finds $\Phi_{\alpha\beta}(t) =  {\rm Tr}(\tau_\alpha \Lambda_t(\tau_\beta))$

\begin{equation}\label{Delta}
  \Lambda_{\alpha\beta}(t) = \left( \begin{array}{c|c} 1 & 0 \\ \hline  \mathbf{x}_t & \Delta_t \end{array} \right) ,
\end{equation}
where $\mathbf{x}_t \in \mathbb{R}^{d^2-1}$, and $\Delta_t$ is a real $(d^2-1)\times (d^2-1)$ matrix. In general a generalised inverse of $\Lambda_{\alpha\beta}(t)$ is not  trace-preserving. {If a generalised inverse fails to preserve trace, then the corresponding propagator also is not trace-preserving.} There is, however, a natural class of trace-preserving generalised inverses corresponding to the following matrix representation

\begin{equation}\label{T-}
  \Lambda_{\alpha\beta}^-(t) = \left( \begin{array}{c|c} 1 & 0 \\ \hline  \mathbf{y}_t & \Gamma_t \end{array} \right)
\end{equation}
{where}, the defining condition $\Lambda(t)\Lambda^-(t)\Lambda(t) =\Lambda(t)$ implies that  $\Gamma_t = \Delta^-_t$, and

$$   \Delta_t(\mathbf{y}_t + \Delta_t^-\mathbf{x}_t) = 0 , $$
that is, $\mathbf{y}_t + \Delta^-_t\mathbf{x}_t \in {\rm Ker}(\Delta_t)$. One finds for the matrix representation of propagator

\begin{equation}\label{}
   \Lambda(t,s):= \Lambda(t)\Lambda^-(s) = \left( \begin{array}{c|c} 1 & 0 \\ \hline  \mathbf{x}_t + \Delta_t \mathbf{y}_s & \Delta_t \Delta_s^- \end{array} \right) .
\end{equation}
This propagator is by construction trace-preserving.

\section{Propagators for Qubit Dynamical Maps}

In the qubit case any quantum channel $\Lambda : M_2(\mathbb{C})(\mathbb{C}) \to M_2(\mathbb{C})(\mathbb{C})$ can be represented as follows \cite{RSW02,QIT,Wolf}

\begin{equation}\label{UV}
  \Lambda = \mathcal{U}\,\Phi\,\mathcal{V}^\dagger ,
\end{equation}
where $\mathcal{U}$ and $\mathcal{V}$ are unitary channels, and $\Phi$ has the following Bloch representation $\Phi_{\alpha\beta} = \frac 12 {\rm Tr}(\sigma_\alpha \Phi(\sigma_\beta))$

\begin{equation}   \label{Tlll}
\Phi_{\alpha\beta} = \begin{pmatrix} 1 & 0 & 0 & 0\\
x_{1} & \lambda_{1} & 0 & 0\\
x_{2} & 0 & \lambda_{2} & 0\\
x_{3} & 0 & 0 & \lambda_{3}
\end{pmatrix},
\end{equation}
where $\lambda_{i}$'s are (up to a sign) the singular values of $\Delta$ as shown in (\ref{Delta}).  Now, since $\mathcal{U}$ and $\mathcal{V}$ are invertible, one has for the generalized inverse of $\Lambda$

\begin{equation}\label{}
  \Lambda^- = \mathcal{V} \Phi^- \mathcal{U}^{\dagger} ,
\end{equation}
that is, the generalized inverse of $\Lambda$ is completely determined by that of $\Phi$. In \cite{CC19} the following theorem was proved

\begin{theorem} There is no CPTP projector $\mathcal{P} : M_2(\mathbb{C}) \to M_2(\mathbb{C})$ projecting $M_2(\mathbb{C})$ to the 3-dimensional subspace of $M_2(\mathbb{C})$.
\end{theorem}
Here we provide an independent proof based on the very concept of generalized inverse. We start with the following:

\begin{Proposition}
	Let $P : \mathcal{H} \to \mathcal{H}$ be a projection onto a subspace $S \subset \mathcal{H}$, and $A : \mathcal{H} \to \mathcal{H}$ be a  linear operator such that  $\rm{Im}(A)=S$. Then, there exists a generalized inverse $A^-$ of $A$ such that $AA^-=P.$
\end{Proposition}
Proof: Let ${\rm dim} \mathcal{H} =n$, and ${\rm dim}S=r<n$. Denote by  $S_0$  the span of the first $r$ of the $n$ vectors in the canonical basis. Let $P_0$ be an arbitrary projection onto this subspace and it has, therefore, the following matrix representation

$$ P_0 = \left( \begin{array}{cc}
\oper_r & X \\
0 & 0
\end{array} \right)\ , $$
where $X$ is an arbitrary  $r \times (n-r)$ matrix. If $A=U\Sigma V^\dagger$ stands for SVD for $A$, then $S_0$ can be transformed to $S$ via $U$, and $P=UP_0U^\dagger$. Now, we choose $P_0=U^\dagger P U$. This fixes the matrix $X$. Hence
$\Sigma=\left( \begin{array}{cc}
D & 0 \\
0 & 0
\end{array} \right)$, where $D$ is an invertible diagonal matrix of dimension $r \times r$. A valid generalized inverse of $\Sigma$ is, $\Sigma^-=\left( \begin{array}{cc}
D^{-1} & D^{-1} X \\
0 & 0
\end{array} \right)$,  and we immediately observe that
\begin{equation}
P_0=\Sigma \Sigma^- .
\end{equation}
Hence, we have $P=AA^-$, where $A^-=V \Sigma^- U^\dagger$ is a generalized inverse of $A$. \hfill $\Box$

Now, consider a projector $\mathcal{P}$ onto 3-dimensional subspace $\Sigma$ in $M_2(\mathbb{C})$ represented via
\begin{equation}\label{}
   \mathcal{P} = \Lambda \Lambda^-  =  \mathcal{U}\,\Phi\, \Phi^-\, \mathcal{U}^\dagger .
\end{equation}
Since the image of $\Lambda$ is 3-dimensional let as assume that in the formula (\ref{Tlll}) one has $\lambda_3=0$ and let us look for the general inverse represented by the general formula (\ref{T-})
\begin{equation}\label{Taaa}
\Phi^-=\begin{pmatrix}1 & 0 & 0 & 0\\
y_{1} & \alpha_{11} & \alpha_{12} & \alpha_{13}\\
y_{2} & \alpha_{21} & \alpha_{22} & \alpha_{23}\\
y_{3} & \alpha_{31} & \alpha_{32} & \alpha_{33}
\end{pmatrix} .
\end{equation}
Using defining property $\Phi\Phi^-\Phi=\Phi$ one finds

$$  \alpha_{11}= \frac 1\lambda_1 \ , \ \ \alpha_{22}= \frac 1\lambda_2 \ , \ \ \alpha_{12} = \alpha_{21} =0 , $$
together with


\begin{equation}\label{}
  y_1 = - \frac{x_1 + \lambda_1 \alpha_{13} x_3 }{\lambda_1} \ , \ \ \ y_2 = - \frac{x_2 + \lambda_2 \alpha_{23} x_3 }{\lambda_2} .
\end{equation}
The remaining parameters are completely free. Hence the Bloch representation of the corresponding projector $\Phi\Phi^-$ reads

\begin{equation}
\Phi\Phi^-=\begin{pmatrix}1 & 0 & 0 & 0\\
- \beta_1 x_3 & 1 & 0 & \beta_1 \\
- \beta_2 x_3 & 0 & 1 & \beta_2 \\
x_3 & 0 & 0 & 0
\end{pmatrix} ,
\end{equation}
with $\beta_1 = \lambda_1\alpha_{13}$ and  $\beta_2 = \lambda_2\alpha_{23}$. One easily finds the corresponding map $\Phi\Phi^-$:

\begin{equation}\label{}
  \Phi\Phi^-(\oper) = \oper - x_3(\beta_1 \sigma_1 + \beta_2 \sigma_2 - \sigma_3) \ ,
\end{equation}
\begin{equation}\label{}
  \Phi\Phi^-(\sigma_3) = \beta_1 \sigma_1 + \beta_2 \sigma_2 ,
\end{equation}
and $\Phi\Phi^-(\sigma_1) =  \sigma_1$, $\Phi\Phi^-(\sigma_2) =  \sigma_2$. To check complete positivity one has to analyze the spectrum of the corresponding Choi matrix

\begin{equation}\label{}
  C = \sum_{i,j=1}^2 E_{ij} \otimes \Phi\Phi^-(E_{ij}) ,
\end{equation}
with $E_{ij}:= |i\rangle \langle j|$. Using

$$   E_{11}= \frac 12 (\oper + \sigma_3) \ , \ \ E_{22} = \frac 12 (\oper - \sigma_3) , \ \ E_{12} = \frac 12 (\sigma_1 + i \sigma_2) , $$
and $E_{21} = E_{12}^\dagger$, one finds

\begin{widetext}
\begin{equation}\label{choi}
  C = \frac 12 \left( \begin{array}{cc|cc} 1+x_3 & (1-x_3)(\beta_1 - i \beta_2) & 0 & 2 \\
   (1-x_3)(\beta_1 + i \beta_2) & 1-x_3 & 0 & 0 \\ \hline
   0 & 0 & 1+x_3 & - (1+x_3) (\beta_1 - i \beta_2) \\ 2 & 0 &  - (1+x_3) (\beta_1 + i \beta_2) & 1-x_3
  \end{array} \right) .
\end{equation}
\end{widetext}
Now, observe that $2 \times 2$ submatrix

$$  \begin{pmatrix} 1+x_3 & 2 \\ 2 & 1-x_3
\end{pmatrix} $$
is not positive, and hence the projector $\Phi\Phi^-$ is not completely positive. \hfill $\Box$

\begin{Remark} Note, that there exists a positive trace-preserving projector onto 3-dimensional subspace of $M_2(\mathbb{C})$. Indeed, to show it recall that all positive maps $\Phi : M_2(\mathbb{C}) \to M_2(\mathbb{C})$ are decomposable \cite{Stormer} which means that the corresponding Choi matrix can be represented as follows

\begin{equation}\label{CCC}
  C = C_1 + C_2^\Gamma ,
\end{equation}
where $C_1,C_2 \geq 0$, and $C_2^\Gamma$ denotes partial transposition. Now, let us observe that (\ref{choi}) satisfies (\ref{CCC}) if and only if $x_3=\beta_1=\beta_2=0$. Indeed, in this case one has

\begin{equation*}\label{}
 2 C = \left( \begin{array}{cc|cc} 1 & 0 & 0 & 1 \\
   0 & 0 & 0 & 0 \\ \hline
   0 & 0 & 0 & 0 \\ 1 & 0 & 0 & 1
  \end{array} \right) + \left( \begin{array}{cc|cc} 0 & 0 & 0 & 0 \\
   0 & 1 & 1 & 0 \\ \hline
   0 & 1 & 1 & 0 \\ 0 & 0 & 0 & 0
  \end{array} \right)^\Gamma .
\end{equation*}
It gives rise to
  \begin{equation}\label{PTP 2D projector}
\Phi\Phi^-=\begin{pmatrix}1 & 0 & 0 & 0\\
0 & 1 & 0 & 0 \\
0 & 0 & 1 & 0 \\
0& 0 & 0 & 0
\end{pmatrix} ,
\end{equation}
and hence unital positive trace-preserving projector $ \mathcal{P} = \mathcal{U}\,\Phi\, \Phi^-\, \mathcal{U}^\dagger$.
{Note,  that (\ref{PTP 2D projector}) is positive due to the fact that it is a unital contraction \cite{Paulsen}.}
\end{Remark}

\vspace{.2cm}

Consider now the structure of projectors onto 2-dimensional subspaces of $M_2(\mathbb{C})$, that is, let us assume that in the formula (\ref{Tlll}) one has $\lambda_2=\lambda_3=0$. Now,  using again the defining property $\Phi\Phi^-\Phi=\Phi$ one finds

$$  \alpha_{11}= \frac 1\lambda_1 \  , \ \  y_1 = - \frac{x_1 + \lambda_1 \alpha_{12} x_2 + \lambda_1 \alpha_{13} x_3 }{\lambda_1} .  $$
The remaining parameters are completely free. Hence the Bloch representation of the corresponding projector $TT^-$ reads

\begin{equation}
\Phi\Phi^-=\begin{pmatrix}1 & 0 & 0 & 0\\
- (\gamma_2 x_2 + \gamma_3 x_3) & 1 & \gamma_2 & \gamma_3 \\
x_2 & 0 & 0 & 0 \\
x_3 & 0 & 0 & 0
\end{pmatrix} ,
\end{equation}
with $\gamma_2 = \lambda_1\alpha_{12}$ and  $\gamma_3 = \lambda_1\alpha_{13}$. One easily finds the corresponding map $\Phi\Phi^-$:

$$   \Phi\Phi^-(\oper) = \oper -  (\gamma_2 x_2 + \gamma_3 x_3)  \sigma_1 + x_2 \sigma_2 + x_3 \sigma_3 , $$
together with

$$    \Phi\Phi^-(\sigma_1)=\sigma_1 \ , \ \  \Phi\Phi^-(\sigma_2) = \gamma_2 \sigma_1 \ , \ \  \Phi\Phi^-(\sigma_3) = \gamma_3 \sigma_1 , $$
and hence the corresponding Choi matrix reads

\begin{widetext}
\begin{equation}\label{C2}
  C = \frac{1}{2} \left(
\begin{array}{cc|cc}
 1+x_3 & \gamma_{3} (1-x_3)-x_2(i+\gamma_{2}) & 0 & 1+i \gamma_{2} \\
  \gamma_{3} (1-x_3)+ x_2(i-\gamma_{2}) & 1-x_3 & 1+i \gamma_{2} & 0 \\ \hline
 0 & 1-i \gamma_{2} & 1+x_3 &  -\gamma_{3} (1+x_3) -x_2(i+\gamma_{2}) \\
 1-i \gamma_{2} & 0 &  -\gamma_{3} (1+x_3) + x_2(i-\gamma_{2}) & 1-x_3 \\
\end{array} \right) .
\end{equation}
\end{widetext}
\begin{Lemma} The Choi matrix (\ref{C2}) is positive semidefinite iff $x_2=x_3=0$ and $\gamma_1=\gamma_2=0$.
\end{Lemma}
Indeed,  let us observe that the following $2 \times 2$ submatrix

$$  \begin{pmatrix} 1+x_3 &  1+i \gamma_{2}  \\  1- i \gamma_{2}  & 1-x_3
\end{pmatrix} $$
is positive iff $1-x_3^2 \geq 1+ \gamma_2^2$ which implies $x_3=\gamma_2=0$. Hence (\ref{C2}) reduces to

\begin{equation}\label{C3}
  C = \frac{1}{2} \left(
\begin{array}{cc|cc}
 1 & \gamma_{3} -i x_2 & 0 & 1 \\
  \gamma_{3} + i x_2  & 1 & 1 & 0 \\ \hline
 0 & 1 & 1 &  -\gamma_{3}-i x_2 \\
 1 & 0 &  -\gamma_{3} + i x_2 & 1 \\
\end{array} \right) .
\end{equation}
The eigenvalues (each with multiplicity 2) of (\ref{C3}) read as follows

$$  1-\sqrt{\gamma_3^2+x_2^2+1},\ 1+ \sqrt{\gamma_3^2+x_2^2+1}, $$
and hence it is evident that they are all non-negative iff $x_2=\gamma_3=0$. \hfill $\Box$

This way we proved that $\Phi\Phi^-$ defines Bloch representation of CPTP projector onto 2-dimensional subspace in $M_2(\mathbb{C})$ if and only if

\begin{equation}
\Phi\Phi^-=\begin{pmatrix}1 & 0 & 0 & 0\\
0 & 1 & 0 & 0 \\
0 & 0 & 0 & 0 \\
0 & 0 & 0 & 0
\end{pmatrix} .
\end{equation}
Note, that the corresponding Kraus representation of $\Phi\Phi^-$ reads $\Phi\Phi^-(X) = \frac 12 (X + \sigma_1 X \sigma_1)$ and hence that of $\Lambda \Lambda^-$ reads as follows

\begin{equation}\label{}
  \Lambda \Lambda^-(X) = \frac 12 (X + U\sigma_1 U^\dagger X U \sigma_1 U^\dagger) ,
\end{equation}
where we used $\mathcal{U}(X) = UXU^\dagger$. Interestingly, {this implies that} any CPTP projector onto 2-dim. subspace is always unital.

\vspace{.2cm}

Finally, if the image is 1-dimensional then the corresponding projector reads

\begin{equation}
\Phi\Phi^-=\begin{pmatrix}1 & 0 & 0 & 0\\
x_1 & 0 & 0 & 0 \\
x_2 & 0 & 0 & 0 \\
x_3 & 0 & 0 & 0
\end{pmatrix} ,
\end{equation}
where the Bloch vector $\mathbf{x}=(x_1,x_2,x_3)$ satisfies $|\mathbf{x}|\leq 1$.

\section{Examples}

In this Section we  illustrate the construction of qubit propagators for divisible dynamical maps using Bloch representation.

\begin{Example} Consider a commutative diagonalizable qubit dynamical map satisfying time-local master equation

\begin{equation}\label{}
  \dot{\Lambda}_t = \mathcal{L}_t \Lambda_t ,
\end{equation}
where
\begin{equation}\label{}
  \mathcal{L}_t =    \gamma_1(t) \mathcal{L}_1 +   \gamma_2(t) \mathcal{L}_2 +  \gamma_3(t) \mathcal{L}_3   ,
\end{equation}
and $\mathcal{L}_k(\rho) = \frac 12 (\sigma_k \rho \sigma_k -\rho)$ (this evolution was already analyzed in \cite{PRL-2018})
The corresponding dynamical map reads

\begin{equation}\label{}
  \Lambda_t(\rho) = \sum_{\alpha=0}^3 p_\alpha(t) \sigma_\alpha \rho \sigma_\alpha ,
\end{equation}
with $\sigma_0 = \oper$, and has the following Bloch representation

\begin{equation}
\Lambda(t)=\begin{pmatrix}1 & 0 & 0 & 0\\
0 & \lambda_1(t) & 0 & 0\\
0 & 0 & \lambda_2(t) & 0\\
0 & 0 & 0 & \lambda_3(t)
\end{pmatrix} ,
\end{equation}
with

$$  \lambda_i(t) = \exp( - \Gamma_j(t) - \Gamma_k(t) ) , $$
where $\{i,j,k\}$ is a permutation of $\{1,2,3\}$, and $\Gamma_k(t) = \int_0^t \gamma_k(\tau)d\tau$. The map $\Lambda_t$ is invertible if all $\Gamma_k(t)$ are finite for finite times. Now, if for example one has $\Gamma_1(t_*)=\infty$, then $\lambda_2(t_*) = \lambda_3(t_*){ =  0}$ which means that the image of $\Lambda_{t_*}$ is $2$-dimensional and of course it is orthogonal to the $2$-dimensional kernel. Now, for any $s > t_*$ (assuming that the image of $\Lambda_s$ is 2-dimensional) one has the following Bloch representation of the corresponding propagator

\begin{equation}
V(t,s)  =\begin{pmatrix}1 & 0 & 0 & 0\\
0 & \lambda_1(t)/\lambda_1(s) & 0 & 0\\
0 & 0 & 0 & 0\\
0 & 0 & 0 & 0
\end{pmatrix} .
\end{equation}
Note that $V_{t,t}$ defines CPTP projector.

\end{Example}

\begin{Example}
Consider the qubit dynamical map given by:
\begin{equation}
\Lambda_t(X) = [1-f(t)] X + f(t) \omega {\rm Tr}(X)
\end{equation}
where $\omega$ is a density matrix, and $f:\mathbb{R}_{\geq 0} \to [0,1]$ is a monotonic function with $f(0)=0$ and $f(t)=1$ for all $t \geq t_*$. It is direct to see that, $\{\Lambda_t\}_{t \geq 0}$ is CP-divisible. In the Pauli basis, with $X=\sum_\alpha x_\alpha \sigma_\alpha $, and $\omega= \frac 12(1 +\sum_k \omega_k \sigma_k)$ the map gives rise to the following Bloch representation
\begin{equation}
\Lambda(t)=\begin{pmatrix}1 & 0 & 0 & 0\\
f(t)\omega_1 & (1-f(t)) & 0 & 0\\
f(t)\omega_2 & 0 & (1-f(t)) & 0\\
f(t)\omega_3 & 0 & 0 & (1-f(t))
\end{pmatrix}
\end{equation}
For $s <t_*$, the map $\Lambda_s$ is invertible, and
\begin{equation}
\Lambda^{-1}(s)=\begin{pmatrix}1 & 0 & 0 & 0\\
-\frac{f(s)\omega_1}{1-f(s)} & \frac{1}{1-f(s)} & 0 & 0\\
-\frac{f(s)\omega_2}{1-f(s)} & 0 & \frac{1}{1-f(s)} & 0\\
-\frac{f(s)\omega_3}{1-f(s)} & 0 & 0 & \frac{1}{1-f(s)}
\end{pmatrix}
\end{equation}
and hence, we have the unique propagator
\begin{equation}
V(t,s) = \begin{pmatrix}1 & 0 & 0 & 0\\
 \omega_1\frac{f(t)-f(s)}{1-f(s)}& \frac{1-f(t)}{1-f(s)} & 0 & 0\\
 \omega_2\frac{f(t)-f(s)}{1-f(s)}& 0 & \frac{1-f(t)}{1-f(s)} & 0\\
 \omega_3\frac{f(t)-f(s)}{1-f(s)}& 0 & 0 & \frac{1-f(t)}{1-f(s)}
\end{pmatrix}
\end{equation}
Evidently, this is a channel for $0\leq s \leq t_*$, and in particular, for $0\leq s \leq t_* \leq t$, it is a CPTP projection operator onto the 1 dimensional subspace spanned by $\omega$. Now, for $0\leq t_* \leq s \leq t$, we have
\begin{equation}
\Lambda(s) = \Lambda(t) =\begin{pmatrix}
1 & 0 & 0 & 0\\
\omega_1 & 0 & 0 & 0\\
\omega_2 & 0 & 0 & 0\\
\omega_3 & 0 & 0 & 0
\end{pmatrix} .
\end{equation}
A direct computation shows that the following matrix is a generalised inverse of $\Lambda_s$ giving rise to the same propagator $V_{t,s}$ through $\Lambda_t \Lambda_s ^-$:
\begin{equation}
\Lambda^-(s) =\begin{pmatrix}
1 & 0 & 0 & 0\\
\alpha_{10} & \alpha_{11} & \alpha_{12} & \alpha_{13}\\
\alpha_{20} & \alpha_{21} & \alpha_{22} & \alpha_{23}\\
\alpha_{30} & \alpha_{31} & \alpha_{32} & \alpha_{33}
\end{pmatrix}
\end{equation}
where all the $\alpha_{ij}$'s are completely arbitrary (this is the gauge freedom of the construction). However, the propagator reads as follows

\begin{equation}
V(t,s) = \Lambda(t)\Lambda^-(s) =\begin{pmatrix}
1 & 0 & 0 & 0\\
\omega_1 & 0 & 0 & 0\\
\omega_2 & 0 & 0 & 0\\
\omega_3 & 0 & 0 & 0
\end{pmatrix} ,
\end{equation}
does not depend on $\alpha_{kl}$.

\end{Example}

Now, we give another example, of a non-diagonalizable, non CP-divisible (but still divisible) qubit dynamical map:

\begin{Example}   \label{EX-NON}
We consider the following Bloch representation of a qubit channel
\begin{equation}
\Psi=\begin{pmatrix}
1 & 0 & 0 & 0\\
0 & 0 & 0 & 1\\
0 & 0 & 0 & 0\\
0 & 0 & 0 & 0
\end{pmatrix} \ .
\end{equation}
One sees that such a matrix is non-diagonalizable, has rank 2, singular values $1,1,0,0,$ eigenvalues $1,0,0,0$, and Kraus rank $2$ with the following Kraus operators
\begin{equation}
K_1=\frac{1}{\sqrt{2}}\begin{pmatrix}
0 & 0\\
-1 & 1
\end{pmatrix}, \
K_2=\frac{1}{\sqrt{2}}\begin{pmatrix}
1 & 1\\
0 & 0
\end{pmatrix} \ ,
\end{equation}
that is, $\Psi(\rho)=\sum_i K_i^\dagger \rho K_i$. One easily checks that the map is trace-preserving, i.e. $\sum_i K_i K_i ^\dagger = \oper$. Now, we define the dynamical map as follows
\begin{equation}
\Lambda_t(X) = [1-f(t)] X + f(t) \Psi(X)
\end{equation}
where $f(t) \in [0,1]$. Suppose, again that $f(t)=1$ for $t \geq t_*$. Clearly, the map is divisible.   Now, for $s < t_*$ one finds for the propagator

\begin{equation}\label{}
 V(t,s) = \Lambda(t) \Lambda^{-1}(s) = \begin{pmatrix}1 & 0 & 0 & 0\\
0 & \frac{1-f(t)}{1-f(s)} & 0 & \frac{f(t)-f(s)}{(1-f(s))^2} \\
0 & 0 & \frac{1-f(t)}{1-f(s)} & 0\\
0 & 0 & 0 & \frac{1-f(t)}{1-f(s)} .
\end{pmatrix}
\end{equation}
Simple analysis of the corresponding Choi matrix shows that in general $V_{t,s}$ is not completely positive and hence $\Lambda_t$ is not CP-divisible. Now, for $t_* \leq s$ one has $\Lambda_{t}=\Lambda_{s}=\Psi$,
and the most general (and trace-preserving) generalised inverse reads
\begin{equation}
\Lambda^-(s) =\begin{pmatrix}
1 & 0 & 0 & 0\\
\alpha_{10} & \alpha_{11} & \alpha_{12} & \alpha_{13}\\
\alpha_{20} & \alpha_{21} & \alpha_{22} & \alpha_{23}\\
0 & 1 & \alpha_{32} & \alpha_{33}
\end{pmatrix}
\end{equation}
where all the $\alpha_{ij}$'s are completely arbitrary, and the corresponding propagator is given by
\begin{equation}
V(t,s)= \Lambda_t \Lambda_{s} ^- =\begin{pmatrix}
1 & 0 & 0 & 0\\
0 & 1 & \alpha_{32} & \alpha_{33}\\
0 & 0 & 0 & 0\\
0 & 0 & 0 & 0
\end{pmatrix} .
\end{equation}
One finds for the Choi matrix

\begin{equation}\label{}
  C = \left(
\begin{array}{cccc}
 1 & \alpha_{33} & 0 & 1-i \alpha_{32} \\
 \alpha_{33} & 1 & 1-i \alpha_{32} & 0 \\
 0 & 1+i \alpha_{32} & 1 & -\alpha_{33} \\
 1+i \alpha_{32} & 0 & -\alpha_{33} & 1 \\
\end{array}
\right) ,
\end{equation}
with the corresponding eigenvalues $1 \pm \sqrt{\alpha_{32}^2+\alpha_{33}^2+1}$. Clearly, $C$ is positive definite if and only if  $\alpha_{32}=\alpha_{33}=0$. Hence, the requirement of complete positivity makes $V_{t,s}$ unique. We stress in this example the image is not complementary to the kernel, and hence one cannot make the choice $\mathcal{C}_s=\rm{Ker}({\Lambda_s})$.
	
\end{Example}

\begin{Example}

Consider now a phase covariant evolution governed by the following time-local master equation

\begin{equation}\label{}
  \dot{\Lambda}_t = \mathcal{L}_t \Lambda_t \ , \ \ \Lambda_0 = {\rm id} ,
\end{equation}
with the following time-local generator \cite{Maniscalco-NJP,Sergey,OSID}
  	
\begin{equation}\label{Maniscalco_exp}
\mathcal{L}_t = \gamma_+(t) \mathcal{L}_+ +   \gamma_-(t) \mathcal{L}_- +  \gamma_3(t) \mathcal{L}_3   ,
\end{equation}
where
	
\begin{eqnarray*}
\mathcal{L}_+(\rho) &=& \frac 12 (\sigma_+ \rho \sigma_- - \frac 12 \{ \sigma_-\sigma_+,\rho \}) , \\
\mathcal{L}_-(\rho) &=& \frac 12 (\sigma_- \rho \sigma_+ - \frac 12 \{ \sigma_+\sigma_-,\rho \}) , \\
\mathcal{L}_3(\rho) &=& \frac 12 (\sigma_z \rho \sigma_z - \rho)  ,
\end{eqnarray*}
with $\sigma_\pm = (\sigma_x \pm i \sigma_y)/2$. In general it defines a non-commutative family of maps, that is, $\mathcal{L}_t \mathcal{L}_s \neq \mathcal{L}_s \mathcal{L}_t$. The corresponding dynamical map $\Lambda_t = \mathcal{T} e^{\int_{0}^{t} \mathcal{L}_\tau d\tau}$ is given by
\begin{equation}
\rho= \left(
\begin{array}{c c}
1-p & \alpha\\
\alpha^* & p
\end{array}
\right) \ \to \ \rho_t = \left(
\begin{array}{c c}
1-p(t) & \alpha(t)\\
\alpha(t)^* & p(t)
\end{array}
\right),
\end{equation}
where
$$  p(t) = e^{-\Gamma(t)}[G(t) + p] \ , \ \ \alpha(t) = \alpha\, e^{-(\Gamma(t)/2+{\Gamma_3}(t))} , $$
with

$$ \Gamma_3(t)=\int_0^t\gamma_3(\tau)d\tau\ ,\ \ \Gamma(t) = \frac{1}{2}\int_0^t(\gamma_+(\tau)+\gamma_-(\tau))d\tau\  , $$
and

$$ G(t) = \frac{1}{2}\int_0^te^{\Gamma(\tau)}\gamma_+(\tau)d\tau \ . $$
One finds the corresponding Bloch representation

\begin{widetext}
\begin{equation}
\Lambda(t) = \begin{pmatrix}
1 & 0 & 0 & 0 \\
0 & e^{-(\Gamma(t)/2+{\Gamma_3}(t))} & 0 & 0 \\
0 & 0 & e^{-(\Gamma(t)/2+{\Gamma_3}(t))} & 0 \\
1-e^{-\Gamma(t)}(2G(t) + 1) & 0 & 0 & e^{-\Gamma(t)}
\end{pmatrix} ,
\end{equation}
\end{widetext}
which already has the Ruskai representation \cite{RSW02}. Now, if  $\Gamma(t) \neq +\infty \neq \Gamma_{3}(t)$, the map is invertible, and hence divisible. For divisibility while non-invertible, we need the following: the rank of the map first decreases from $4$ to $2$  at $t=t_{1}$, and $\Gamma_{3}(t)=+\infty$ for $t\geq t_{1}$. Hence, for $s\geq t_{1}$, we have
\begin{equation}
\Lambda(s) = \begin{pmatrix}
1 & 0 & 0 & 0 \\
0 & 0 & 0 & 0 \\
0 & 0 & 0 & 0 \\
1-e^{-\Gamma(s)}(2G(s) + 1) & 0 & 0 & e^{-\Gamma(s)}
\end{pmatrix} .
\end{equation}
The most general trace-preserving generalized inverse of this map can be written as follows

\begin{equation}
\Lambda^-(s) = \begin{pmatrix}
1 & 0 & 0 & 0 \\
y_1 & \alpha_{11} & \alpha_{12} & \alpha_{13} \\
y_2 & \alpha_{21} & \alpha_{22} & \alpha_{23} \\
-e^{\Gamma(s)}+2G(s) + 1 & \alpha_{31} & \alpha_{32} & e^{\Gamma(s)}
\end{pmatrix}
\end{equation}
where all the $y_{i}$'s and $\alpha_{ij}$'s are arbitrary. Hence, the corresponding propagator, for $t\geq s\geq t_{1} > 0$ can be written as:
\begin{equation}
V(t,s) = \begin{pmatrix}
1 & 0 & 0 & 0 \\
0 & 0 & 0 & 0 \\
0 & 0 & 0 & 0 \\
\xi(t,s) & \alpha_{31}e^{-\Gamma(t)} & \alpha_{32}e^{-\Gamma(t)} & e^{-(\Gamma(t)-\Gamma(s))}
\end{pmatrix}\ ,
\end{equation}
where

\begin{eqnarray*}
   \xi(t,s) &=& (1-e^{-\Gamma(t)}(2G(t) + 1)) \\
   &-& e^{-(\Gamma(t)-\Gamma(s))}(1-e^{-\Gamma(s)}(2G(s) + 1)) \ .
\end{eqnarray*}
One finds for the corresponding projection
\begin{equation}
V(s,s)= \begin{pmatrix}
1 & 0 & 0 & 0 \\
0 & 0 & 0 & 0 \\
0 & 0 & 0 & 0 \\
0 & \alpha_{31}e^{-\Gamma(s)} & \alpha_{32}e^{-\Gamma(s)} & 1
\end{pmatrix}
\end{equation}
The Choi matrix of this map reveals that it is CP iff $\alpha_{31}=\alpha_{32}=0$, which is, when it is just {complete decoherence in the $z$-basis}. If the rank decreases once more at $t=t_{2}\geq t_{1}$, that is, if $\Gamma(t)=+\infty$ for all $t\geq t_2$, then  for $t\geq t_2$
\begin{equation}
\Lambda_t = \begin{pmatrix}
1 & 0 & 0 & 0 \\
0 & 0 & 0 & 0 \\
0 & 0 & 0 & 0 \\
1 & 0 & 0 & 0
\end{pmatrix} ,
\end{equation}
which is a CPTP projector onto the space spanned by vacuum state  $|0\rangle \langle 0|$.

\end{Example}

\section{Conclusion}

Divisibility of dynamical maps provides an important characterization of quantum evolution. In particular CP-divisible maps are often consider as a mathematical representation of quantum Markovian evolution \cite{Angel, BOGNA, PRL-2018, CC19, datta}. For invertible maps both P- and CP-divisible maps are fully characterised by the monotonicity of the trace distance $\| \Lambda_t(X)\|_1$ and $\| ({\rm id}\otimes  \Lambda_t)(X)\|_1$ for all Hermitian operators in $\mathcal{H}$ and $\mathcal{H} \otimes \mathcal{H}$, respectively. However, for maps that are not invertible the problem is still open \cite{Angel}. In particular if the map is not invertible one can not use the standard construction of the propagator $\Lambda_t \Lambda_s^{-1}$. In this paper we proposed a natural generalization, that is, we replaced the inverse by a generalized inverse $\Lambda_s^-$. Such construction is perfectly consistent with local composition law, that is, $V_{t,s} = V_{t,u} V_{u,s}$ for $t \geq u \geq s$. Clearly, this construction is highly non-unique since $\Lambda_s^-$ is not uniquely defined. Interestingly, in all the examples we studied, it was found that the CPTP propagator {constructed from a generalized inverse} is unique though the map is non-invertible, which is an interesting observation perhaps not predictable through earlier works in this direction.

It turns out that the technique of generalized inverses provides a very effective tool for attacking such kind of problems. For example, it was shown in \cite{CC19} that there is no qubit channel which is also a projection onto a 3-dimensional subspace. In this paper this interesting result was derived in a natural way via the use of general inverses for qubit maps. It would be interesting to link the concept of generalized inverse to `time reversal' operation considered e.g. in \cite{TimeRev}


\begin{thebibliography}{99}	

\bibitem{Open1} H.-P. Breuer and F. Petruccione, The Theory of Open
Quantum Systems (Oxford University Press, Oxford, 2007).

\bibitem{Open2} A. Rivas and S. F. Huelga, Open Quantum Systems. An
Introduction (Springer, Heidelberg, 2011).

\bibitem{QIT} M. A. Nielsen and I. L. Chuang, {\em Quantum Computation and Quantum Information},
(Cambridge University Press, Cambridge, 2010).

\bibitem{Paulsen} V. Paulsen, {\em Completely Bounded Maps and Operator
Algebras} (Cambridge University Press, Cambridge, 2003).

\bibitem{Stormer} E. St{\o}rmer, Positive Linear Maps of Operator Algebras,
Springer Monographs in Mathematics (Springer, New York,
2013).

\bibitem{Wolf} M. M. Wolf, {\em Quantum Channels \& Operations: Guided Tour},  URL: https://wwwm5.ma.tum.de/foswiki/pub/M5/Allgemeines\\ /MichaelWolf/QChannelLecture.pdf.

\bibitem{GKS} V. Gorini, A. Kossakowski, E.~C.~G. Sudarshan, J. Math. Phys. \textbf{17}, 821 (1976).

\bibitem{L} G. Lindblad, Comm. Math. Phys. \textbf{48}, 119 (1976).

\bibitem{NM1} \'A. Rivas, S. F. Huelga, and M. B. Plenio, Rep. Prog. Phys.
{\bf 77}, 094001 (2014).

\bibitem{NM2} H.-P. Breuer, E.-M. Laine, J. Piilo, and B. Vacchini, Rev.
Mod. Phys. {\bf 88}, 021002 (2016).

\bibitem{NM3} I. de Vega and D. Alonso, Rev. Mod. Phys. {\bf 89}, 015001
(2017).

\bibitem{NM4} L. Li, M. J.W. Hall, and H. M. Wiseman, Phys. Rep. {\bf 759}, 1
(2018).

\bibitem{Piilo}  C.F Li, G.C Guo, and J Piilo, EPL (Europhysics Letters) {\bf 127} (5), 50001; EPL (Europhysics Letters) {\bf 128} (3), 30001.

\bibitem{RHP} \'A. Rivas, S.F. Huelga, and M.B. Plenio, Phys. Rev. Lett. {\bf 105}, 050403
(2010).

\bibitem{BLP} H.-P. Breuer, E.-M. Laine, and J. Piilo, Phys. Rev. Lett. {\bf 103},
210401 (2009).

\bibitem{Wolf-1} M. M. Wolf, J. Eisert, T. S. Cubitt, and J. I. Cirac, Phys. Rev. Lett. {\bf 101}, 150402 (2008).

\bibitem{Wolf-2} M. M. Wolf and J. I. Cirac, Comm. Math. Phys., {\bf 279}, 147 (2008).

\bibitem{Mario} D. Davalos, M. Ziman, and C.  Pineda, Quantum {\bf 3}, 144 (2019).

\bibitem{Monachium} M. C. Caro and B. Graswald, Necessary Criteria for Markovian Divisibility of Linear Maps,  arXiv:2009.06666.

\bibitem{Fabio} F. Benatti, D. Chru\'sci\'nski, and S. Filippov,  Phys. Rev. A. {\bf 95},  012112 (2017).

\bibitem{Sabrina} D. Chru\'sci\'nski and S. Maniscalco, Phys. Rev. Lett. {\bf 112}, 1204 (2014).

\bibitem{Farrukh} D. Chru\'sci\'nski and  F. Mukhamedov, Phys. Rev. A. {\bf 100},  052120 (2019).

\bibitem{Sergey-2017} S. N. Filippov, J. Piilo, S. Maniscalco, and M. Ziman, Phys. Rev. A {\bf 96}, 032111 (2017).

\bibitem{Junu-DC} J. Bae and D.  Chru\'sci\'nski, Phys. Rev. Lett. {\bf 117}, 050403 (2016).

\bibitem{Chem} V. Reimer, M.R. Wegewijs, K. Nestmann, and M. Pletyukhov,  J. Chem. Phys. {\bf 151}, 044101 (2019).

\bibitem{Angel} D. Chru\'sci\'nski, A. Kossakowski, and \'A. Rivas,  Phys. Rev. A {\bf 83}, 052128 (2011).

\bibitem{Janek} J. Ko{\l}ody\'nski, S.  Rana, and A.  Streltsov, Phys. Rev. A {\bf 101}, 020303(R) (2020).

\bibitem{Acin-cor}  D. De Santis, M. Johansson, B.  Bylicka, N.K. Bernardes, and A. Ac\'in, Witnessing non-Markovian dynamics through correlations, arXiv:1903.12218

\bibitem{Johansson} D. De Santis and M. Johansson,  New J. Phys. {\bf 22}, 093034 (2020)

\bibitem{Modi-1}  S. Milz, M. S. Kim, F.A. Pollock, and K. Modi, Phys. Rev. Lett. {\bf 123}, 040401 (2019).

\bibitem{Modi-2} F.A. Pollock, C. Rodr\'iguez-Rosario, T. Frauenheim, M. Paternostro, and K.  Modi, 	Phys. Rev. Lett. 120, 040405 (2018).

\bibitem{Cresser1} E. Andersson, J. D. Cresser, and M. J. W. Hall, J. Mod. Opt. {\bf 54}, 1695 (2007).

\bibitem{Cresser2} J.D. Cresser and C. Facer, Optics Communications {\bf 283}, 773 (2010).

\bibitem{BOGNA} B. Bylicka, M. Johansson, and A. Ac\'in, Phys. Rev. Lett. {\bf 118}, 120501 (2017).

\bibitem{PRL-2018} D. Chru\'{s}ci\'{n}ski, \'A. Rivas, and E. St{\o}rmer, Phys. Rev. Lett. {\bf 121}, 080407 (2018).

\bibitem{CC19} S. Chakraborty and D. Chru\'{s}ci\'{n}ski, Phys. Rev. A {\bf 99}, 042105 (2019)

\bibitem{datta} F. Buscemi and N. Datta, Phys. Rev. A {\bf 93}, 012101 (2016).


\bibitem{HEL} C.W. Helstrom, {\em Quantum Detection and Estimation Theory},
(Academic Press, New York, 1976).

\bibitem{JUNU} J. Bae and L.-Ch. Kwek, J. Phys. A: Math. Theor. {\bf 48},  083001 (2015)

\bibitem{Kumar} C. R. Rao and S. K. Mitra, {\em Generalised Inverse of a Matrix and its Applications},
 New York: John Wiley \& Sons, 1971

\bibitem{GI} A. Ben-Israel and N.E.  Thomas, {\em Generalized inverses: Theory and applications},  (New York, NY: Springer, 2003).

\bibitem{Horn} R.A. Horn and C.R. Johnson,  {\em Matrix Analysis}, (Cambridge University Press, 1985)

\bibitem{Yanai} H. Yanai, K. Takeuchi, and Y. Takane, {\em Projection Matrices, Generalized Inverse Matrices, and Singular Value Decomposition},
    (Springer, 2011)

\bibitem{Watrous} J. Watrous, {\em The Theory of Quantum Information}, (Cambridge University Press, 2018)

\bibitem{Gilchrist} {A. Gilchrist, D. R. Terno, C. J. Wood, {\em Vectorization of quantum operations and its use}, arXiv:0911.2539v2}


\bibitem{Kato} T. Kato, {\em Perturbation Theory for Linear Operators}, 2nd ed. (Springer, Berlin, 1980).

\bibitem{RSW02} M.B. Ruskai, S. Szarek, and E. Werner, Lin. Alg. Appl. {\bf 347}, 159  (2002).

\bibitem{Maniscalco-NJP} J. Teittinen, H. Lyyra, B. Sokolov, and S. Maniscalco, New Journal of Physics {\bf 20}, 073012 (2018).

\bibitem{Sergey} S. N. Filippov, A. N. Glinov, and L. Lepp\"aj\"arvi,  Lobachevskii J. Math. {\bf 41}, 617-630 (2020). (available as arXiv:1911.09468).

\bibitem{OSID} D. Chru\'{s}ci\'{n}ski, Open Syst. Inf. Dyn. {\bf 21}, 1440004 (2014).

\bibitem{TimeRev} E. Aurell, J. Zakrzewski and K. {\.{Z}}yczkowski, J. Phys. A: Math. Theor. {\bf 48}, 38FT01 (2015).

\end{thebibliography}


\section*{Acknowledgements}  DC was supported by the National Science Centre project 2018/30/A/ST2/00837. UC would like to thank the TAPS programme (2019) of the Faculty of Physics, Astronomy and Informatics of the Nicolaus Copernicus University in Toru\'{n} for his stay in Poland in summer 2019.

\end{document}